\documentclass[twocolumn,showpacs,preprintnumbers,amsmath,amssymb]{revtex4-1}

\usepackage{graphicx}
\usepackage{amssymb}
\usepackage{hyperref}

\begin{document}

\title{Physical properties of the very heavy fermion YbCu$_4$Ni}

\author{J.G. Sereni$^1$, I. \v{C}url\'ik $^2$, M. Giovannini$^3$, A. Strydom$^{4,5}$, M. Reiffers$^2$}
\address{$^1$Department of Physics, CAB-CNEA, CONICET, 8400 San
Carlos de Bariloche, Argentina\\
$^2$Faculty of Humanities and Natural Sciences, University of
Pre\v{s}ov, 17. novembra 1, Pre\v{s}ov, Slovakia\\
$^3$Department of Physics and CNR-SPIN, University of Genova, Via
Dodecaneso
31, Genova, Italy\\
$^4$Highly Correlated Matter Research Group, Physics Department,
University of Johannesburg, PO Box 524, Auckland Park 2006, South
Africa\\
$^5$Max Planck Institute for Chemical Physics of Solids,
Nöthnitzerstr.40, D-01187 Dresden, Germany.}

\begin{abstract}

{The physical properties of the very heavy fermion YbCu$_4$Ni were
characterized through structural, magnetic, thermal and transport
studies along nearly four decades of temperature ranging between
50\,milikelvin and 300\,K. At high temperature, the crystal
electric field levels splitting was determined with $\Delta_1
(\Gamma_6)= 85$\,K and $\Delta_2 (\Gamma_8) \approx 200$\,K, the
latter being a quartet in this cubic symmetry. An effective
magnetic moment $\mu_{eff} \approx 3\mu_B$ is evaluated for the
$\Gamma_7$ ground state, while at high temperature the value for a
Yb$^{3+}$ ion is observed. At low temperature this compounds shows
the typical behavior of a magnetically frustrated system
undergoing a change of regime at a characteristic temperature
$T^*\approx 200$\,mK into a sort of Fermi-liquid type 'plateau'of
the specific heat: $C_m/T|_{T\to 0}$ = const. The change in the
temperature dependence of the specific heat coincides with a
maximum and a discontinuity in respective inductive and
dissipative components of the ac-susceptibility. More details from
the nature of this ground state are revealed by the specific heat
behavior under applied magnetic field.}

\end{abstract}
\date{\today}

\maketitle

\section{Introduction}

There is an increasing number of new Yb-based compounds which do
not show long range magnetic order down to the milikelvin range of
temperature. Since they are characterized by having robust
magnetic moments and by exhibiting a divergent increase of
magnetic correlations in their paramagnetic phase upon cooling
\cite{YbCu5-xAux}, it is expected that they belong to a different
class of materials than those placed around a quantum critical
point \cite{QPT07}. The strong increase of the density of magnetic
excitations is reflected in power law dependencies of the specific
heat: $C_m(T)/T\propto 1/T^Q$, with exponents ranging between $1<
Q \leq 2$ that clearly exceed the $C_m/T|_{T\to 0}$ values of
usual non-fermi-liquids (NFL) \cite{Stewart01} that increase as a
$-ln(T/T_0)$ with decreasing temperature. These compounds can be
called very-heavy fermions (VHF) because their $C_m/T|_{T\to 0}$
range between $\approx 5$ and $\approx 12$\,J/molK$^2$
\cite{JLTP18}.

Two scenarios were proposed for the lack of long range magnetic
order in these compounds, one due to a very weak magnetic
interaction between Yb ions and the other to magnetic frustration.
The former applies to YbPt$_2$In and YbPt$_2$Sn compounds that
show the record high $\approx 12$\,J/molK$^2$ value \cite{YbPt2Sn}
after undergoing a maximum of $C_m(T)/T$ around 250\,mK. Magnetic
frustration can be originated by peculiar geometric coordination
of the magnetic moments \cite{Ramirez06} or by the competition
between magnetic interactions of different nature. Triangular (2D)
or tetrahedral (3D) are the typical atomic configurations
producing frustration at low temperature, with the striking
feature that a number of Yb systems show a coincident plateau of
$C_m/T|_{T\to 0}= 7\pm 0.5$\,J/molK$^2$ below a deviation from
their power law dependence of $C_m/T(T)$ within the mK range
\cite{JLTP18} at a characteristic temperature $T^*$.

The mentioned absence of magnetic order in these VHF was verified
by different type of measurements besides specific heat. For
example, electrical resistivity ($\rho$) of YbCu$_{5-x}$Au$_x$
\cite{YbCu5-xAux} and PrInAg$_2$ \cite{PrInAg2} decreases
continuously with temperature around $T^*$ without showing any
discontinuity in their thermal slopes after undergoing a broad
maximum. The lack of magnetic order in YbCu$_{5-x}$Au$_x$ down to
0.02\,K was confirmed through NQR and $\mu$SR measurements
\cite{Caretta}.

A relevant aspect of the VHF compounds concerns their potential
application for adiabatic demagnetization refrigeration based upon
the large amount of entropy accumulated below 1\,K that can be
removed under magnetic field. This property was properly explored
in at least two compounds: YbPt$_2$In \cite{GrunerNat} and
(Yb$_{1-x}$Sc$_x$)Co$_2$Zn$_{20}$ \cite{Gegenw}. The latter
YbCo$_2$Zn$_{20}$ \cite{YbCo2Zn20} together with YbBiPt
\cite{YbBiPt} belong to the mentioned class of compounds showing
similar $C_m/T|_{T\to 0}\approx 7$\,J/molK$^2$ value below $T^*$.
After having recognized the YbCu$_{5-x}$Au$_x$ ($0.4\leq x \leq
0.7$) family as belonging to this "plateau" type group while the
stoichiometric compound YbCu$_{4}$Au presents a cusp at $T\approx
0.8$\,K \cite{YbCu4Au}, we have searched for an alternative
composition in order to identify the main property producing such
a different behavior. With this aim we have studied the physical
properties of the isotypic compound YbCu$_4$Ni, where nearly
isochoric Ni atoms with $\approx 3 \%$\, smaller metallic radius
than Cu, instead of Au that is $\approx 20\%$\ larger.

\section{Experimental details}

The polycrystalline samples of YbCu$_4$Ni and LuCu$_4$Ni were
prepared after weighing stoichiometric amounts of the elements
with the following nominal purity Yb-99.9\% mass, Lu-99.9\% mass,
Cu-99.999\% mass and Ni-99,995\% mass. Then the elements were
enclosed in small tantalum crucibles and sealed by arc welding
under pure argon. The samples were melted in an induction furnace
under a stream of pure argon. To ensure homogeneity during the
melting, the crucibles were continuously shaken. The samples were
then annealed in a resistive furnace for two weeks at 700 °C and
finally quenched in cold water.

\begin{figure}[tb]
\begin{center}
\includegraphics[width=20pc]{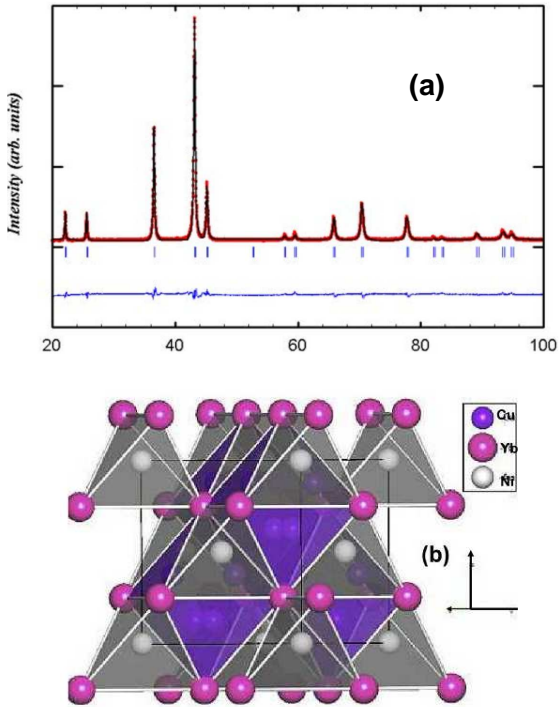}
\end{center}
\caption{a) The experimental x-ray diffraction powder pattern of
YbCu$_4$Ni compared with the calculated diffraction diagram. The
experimental data are shown by symbols, whereas the line through
the data represents the results of the Rietveld refinement. The
lower curve is the difference curve. The ticks indicate Bragg peak
positions. b) Crystal structure seen as edge-sharing tetrahedra
with Yb ions located at the vertices (unit cell indicated by black
lines) after \cite{Giova15}.} \label{FF1}
\end{figure}

The two samples YbCu$_4$Ni and LuCu$_4$Ni were characterized by
optical and electron microscopy and by quantitative electron probe
microanalysis. The crystal structure was examined by x-ray
diffraction (XRD) using a K{$\alpha$} radiation. Rietveld matrix
full profile refinements were performed using the program FULLPROF
\cite{Fullprof}.

Specific heat measurements on YbCu$_4$Ni were carried out in a
commercial PPMS equipment within the temperature range $0.4 < T <
300$\,K, in zero and applied magnetic fields up to 9\,T. Using the
same device the temperature dependence of electrical resistivity
was measured within the mentioned ranges of temperature and field
for magnetoresistance. The low temperature resistivity
measurements (50\,mK $< T <$ 4\,K) were performed using a
four-coils configuration, applying an AC current of 0.2\,mA with a
$f=18.3$\,Hz frequency. The temperature dependence of
magnetization and magnetic susceptibility were obtained using
commercial MPMS. At low temperature (between 50\,mK and 4\,K)
AC-magnetic susceptibility was measured applying an excitation
intensity of 1\,Oe using different frequencies: $f=4$, 1 and
0.1\,kHz.

\section{Experimental Results}
\subsection{Crystal structure of YbCu$_4$Ni}

Single-phase sample material of YbCu$_4$Ni was obtained showing a
face-centered cubic (FCC) structure of MgCu$_4$Sn type (ternary
derivative of the AuBe$_5$) with lattice constant: $a = 6.943
\AA$. The structural results of the Rietveld profile fitting are
summarized in Table I and the corresponding diffraction pattern is
shown in Fig.~\ref{FF1}a. The structure was checked with respect
to ordered or disordered locations of atoms in the structure, but
the compound resulted to be a fully ordered compound with a full
occupation of the atoms in each site. As depicted in
Fig.~\ref{FF1}b, this FCC lattice can be viewed as a network of
edge-sharing tetrahedra with Yb magnetic ions located at the
vertices \cite{Giova15} being a 3D analogue of a triangular
lattice.  The details of the structural refinement for the
isotypic LuCu$_4$Ni compound were reported in a previous work
\cite{12Curlik}.

\begin{table}
\centering{ \caption{Structural parameters of YbCu$_4$Ni refined
from x-ray data in the space group $F-43m$. Residual values: $R_B
= 4.10$\%, $R_F = 5.58$\%, $R_{WP} = 6.86$\%.}
\medskip
\begin{tabular}{llllc}
\\
\hline \hline
 Atom   .&   Site    .&   $x$   &   $y$    &   $z$ \\
\hline \hline
Yb   & 4c  & 0.25 & 0.25 & 0.25 \\
\hline
Cu   & 16c &  0.6259(1)  &  0.6259(1)  &   0.6259(1)  \\
\hline
Ni   & 4c  & 0 & 0 & 0 \\
\hline \hline
\end{tabular}}
\end{table}

\subsection{Magnetic susceptibility and magnetization}

\begin{figure}[tb]
\begin{center}
\includegraphics[width=19pc]{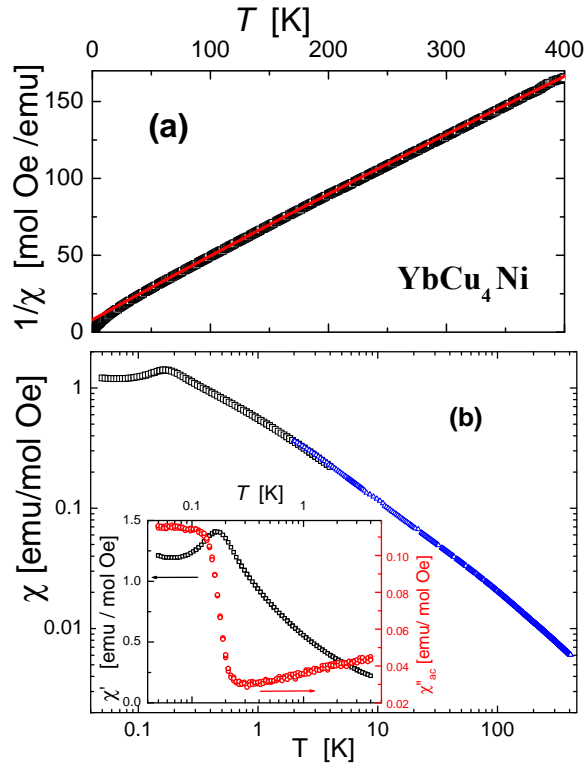}
\end{center}
\caption{a) High temperature dependence of the inverse magnetic
susceptibility, b) low temperature inductive component of
AC-susceptibility scaled with high temperature DC-susceptibility
in a double logarithmic representation. Inset: inductive ($\chi'$,
left axis) and dissipative ($\chi"$, right axis) components of AC
susceptibility in a semi-logarithmic representation.} \label{FF2}
\end{figure}

In Fig.~\ref{FF2}a the inverse magnetic susceptibility ($1/\chi$)
of YbCu$_4$Ni is shown between 2 and 400\,K. This temperature
dependence can be described by a simple Curie-Weiss law with a
negligible Pauli-type contribution ($\chi_0=4\times10^{-4}$
emu/mol\,Oe): $\chi(T)=C_c/(T+\theta_P) + \chi_0$. The Curie
constant ($C_c$) extracted from a fit on the high temperature
range: 70 - 300\,K, reveals an effective magnetic moment
$\mu_{eff} = 4.42 \mu_B$, close to the free Yb$^{3+}$ ion value.
The paramagnetic Curie-Weiss temperature $\theta_P= -19$\,K
indicates a dominating antiferromagnetic exchange at high
temperature and a weak Kondo effect affecting the excited crystal
electric field (CEF) levels. Below $T=70$\,K, $1/\chi$ turns down
due to a smaller $\mu_{eff}$ value of the ground state (GS), which
between $5\geq T \geq 2$\,K is evaluated as $\mu^{GS}_{eff}\approx
3.1 \mu_B$ with a weake intensity of the molecular field
$\theta^{GS}_P = -1.3$\,K.

Below 4\,K the magnetic susceptibility was measured by an AC-
method down to 50\,mK, without detecting a frequency dependence,
see the inset in Fig.~\ref{FF2}b. The results for $f=4$\,kHz show
a maximum in the inductive component $\chi'(T)$ at $T^{\chi}_{max}
\approx 170$\,mK and a clear step in the dissipative component
($\chi"$) at the same temperature. This feature will be discussed
in the following section together with the onset of coherence in
$\rho(T)$ and the $C_m(T)/T|_{T\to 0}$ plateau. The low
temperature $\chi'(T)$ dependence is overlapped with the high
temperature $\chi_{DC}$ in Fig.~\ref{FF2}b within 2 to 4\,K, using
the coincident thermal slope as a matching criterion, notice the
double logarithmic representation.

\begin{figure}[tb]
\begin{center}
\includegraphics[width=19pc]{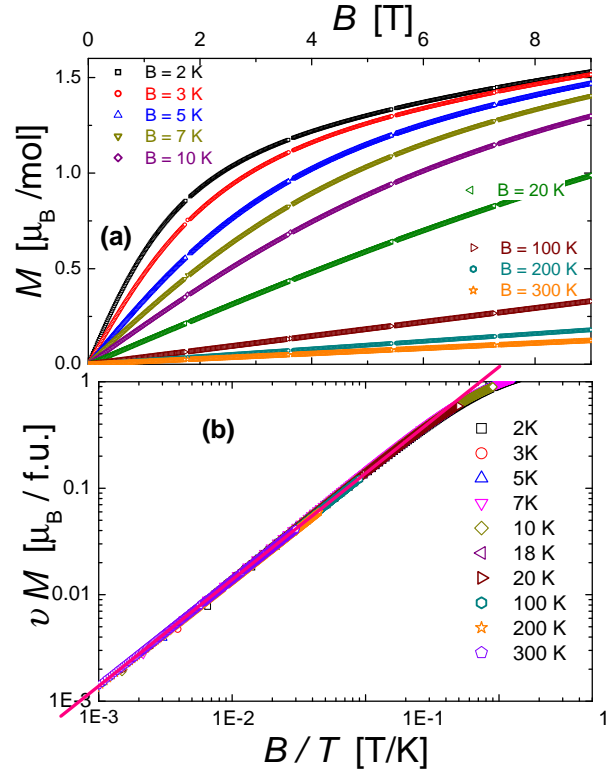}
\end{center}
\caption{(a) Scaled magnetization measurements versus field up to
9\,T performed between 2\,K and room temperature. (b) Collapsed
curves scaling with $x=B/T$ and $y=\nu \times M$, with $\nu$
accounting for the CEF progressive thermal occupation. Notice the
double logarithmic scale.} \label{FF3}
\end{figure}

The magnetization measurements performed between 2\,K and room
temperature applying magnetic field up to 9\,T are presented in
Fig.~\ref{FF3}a. Since this system does not show magnetic order it
is interesting to recognize up to what extent it can be considered
a standard paramagnet (e.g. Curie-Weiss type) at low temperature.
The strong curvature observed in Fig.~\ref{FF3}a at $T=2$\,K
raises the question whether it is due to the effect of an emerging
very low temperature interaction or it corresponds to the typical
curvature of a Brillouin function $B_J(x)$, with $x=B/T$. For this
purpose we have scaled all $M(B)$ curves between $2\leq T \leq
300$\,K using $x=B/T$ as abscissa and normalizing the
magnetization as $M/\nu$, where $\nu$ accounts for progressive
thermal (i.e. Boltzmann) occupation of the excited CEF levels. It
can be appreciated in Fig.~\ref{FF3}b how all curves collapse into
a unique curve once these two scaling parameters, which
characterize the evolution of a typical Curie-Weiss paramagnetic
system, are applied.

\subsection{Electrical resistivity and magnetoresistence}

\begin{figure}[tb]
\begin{center}
\includegraphics[width=19pc]{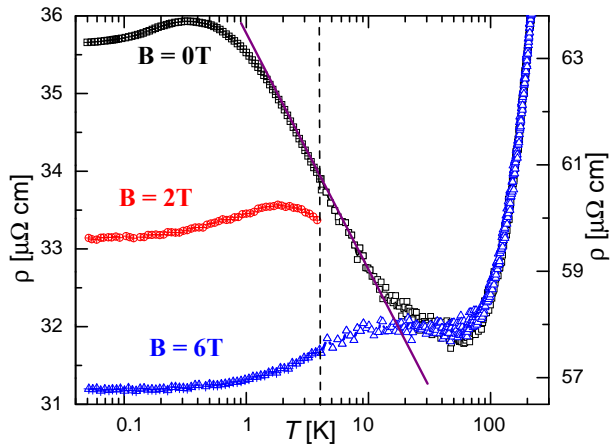}
\end{center}
\caption{Logarithmic temperature dependence of the electrical
resistivity within four decades of temperature at $B=0$ and 6\,T,
measured in different cryostats. Dashed (vertical) line indicates
the data matched at $T=4$\,K using a ratio of 1.8 in the form
factor (see the text). Straight (purple) line remarks the
logarithmic $T$ dependence between $2< T < 10$\,K. The $\rho(T\leq
4,B=2T)$\,K data are included to show the $T^{\rho}_{max}$
variation with field.} \label{FF4}
\end{figure}

The temperature dependence of the electrical resistivity $\rho(T)$
of YbCu$_4$Ni is presented in Fig.~\ref{FF4} within four decades
of temperature at zero and $B=6$\,T, taking the high temperature
data from \cite{rhoActa}. Since these results were obtained in
different cryostats and from different pieces of the
poly-crystalline sample, a ratio 1.8 between respective form
factors is observed (compare left low temperature with right high
temperature axis in Fig.~\ref{FF4}). Such a difference is
attributed to a different electrical connectivity between grains
within the sample. Between room temperature and $T\approx 60$\,K,
$\rho(T)$ decreases mostly due to the decreasing phonon
contribution. Then, below $T\approx 10$\,K, $\rho(T)$ shows a
typical $\rho(T)\propto - ln(T/T_0)$ Kondo increase for the $B$=0
data.

After undergoing a broad maximum, centered at $T^{\rho}_{max}
\approx 350$\,mK in the B=0 data, $\rho_0|_{T\to 0}$ flattens.
Magnetic field induces negative magnetoresistance at low
temperature while the $T^{\rho}_{max}$ is shifted to higher
temperature. Low temperature $\rho(T\leq 4)$\,K data, measured
with $B=2$\,T, are included to confirm this field dependent
tendency which seems to be proportional to $B$.

\subsection{Magnetic contribution to the specific heat}

\begin{figure}[tb]
\begin{center}
\includegraphics[width=19pc]{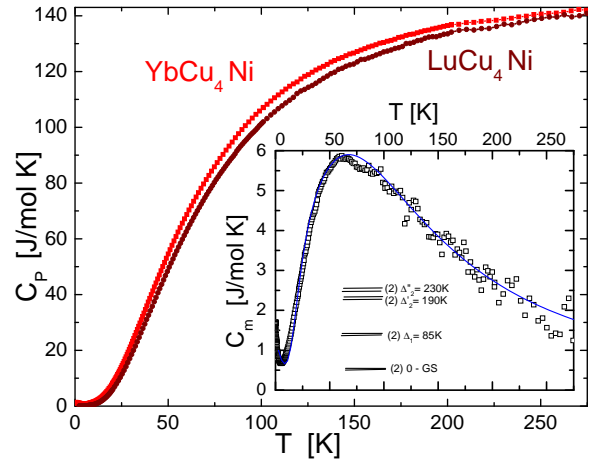}
\end{center}
\caption{Specific heat temperature dependencies up to room
temperature of YbCu$_4$Ni and LuCu$_4$Ni. Inset: high temperature
magnetic contribution and a fit accounting for the CEF levels and
GS contributions.} \label{FF5}
\end{figure}

Fig.~\ref{FF5} shows the high temperature dependence of the
YbCu$_4$Ni specific heat. Since measured specific heat is usually
dominated by electron band ($C_{el}$), phonon ($C_{ph}$) and
magnetic ($C_{m}$) contributions $C_P = C_{el} + C_{ph}+C_{m}$. In
order to subtract the non magnetic components: $C_{el} + C_{ph} =
\gamma T+ \beta T^3$, we have investigated also the LuCu$_4$Ni
isotype. Its specific heat was also measured within the $0.4 \leq
T \leq 300$\,K and is included in Fig.~\ref{FF5}. Both compounds
are close to reach the expected Dulong-Petit values at high
temperature. At low temperature, the extracted electron Sommerfeld
coefficient for LuCu$_4$Ni is: $\gamma$(Lu)=9.5\,mJ/mol K$^2$, a
typical value for non-magnetic lanthanides compounds. Its Debye
temperature: $\theta_D = 325$\,K, was computed from the $\beta
\approx 0.5$\,mJ/molK$^4$ coefficient.

At the low temperature limit (i.e. $T<0.2$\,K), the specific heat
measurements of YbCu$_4$Ni show a clear increase attributed to
nuclear contribution ($C_n$). Since this increase is found to
proceed as $C_n=A_n/T^2$, in order to compute its contribution we
have plot the measured data using a $C_P/T=A_n/T^3+C_m(T)/T$
temperature dependence. From the fit performed on the studied
sample below $T=0.22$ K one extracts
$A_n=1.35\cdot10^{-3}$\,J\,K/mol and $C_m/T|_{T \to 0} = 7.5$
J/molK$^2$. This $A_n$ coefficient is very similar to the one
obtained for YbCu$_{5-x}$Au$_x$ \cite{YbCu5-xAux} and the value:
$C_m/T|_{T \to 0} = 7.5$ J/molK$^2$, places YbCu$_4$Ni into the
class of VHF showing a 'plateau' at $T\to 0$ \cite{JLTP18}. No
specific heat jump is observed in $C_m(T)$ rather a well defined
cusp at $T^{C_P}_{max} \approx 270$\,mK as depicted in
Fig.~\ref{FF8}.

\section{Discussion}

\subsection{Crystal electric field effect}

In the inset of Fig.~\ref{FF5} one can see that $C_m(T)$ starts to
increase around 10\,K once the GS contribution is overcome by that
of the excited CEF levels. This Schottky type anomaly, centered at
$\approx 60$\,K, is attributed to the Yb excited crystal electric
field CEF levels. In a cubic symmetry, the 8-fold ($N=2J+1=8$)
degenerate ground state given by Hund's rule for a $J=7/2$ angular
moment splits into two doublets ($\Gamma_6$ and $\Gamma_7$) and a
quartet ($\Gamma_8$) according to these calculations of Lea, Leask
and Wolf (LLW) \cite{LLW}. In order to extract the level spectrum,
we have fitted the experimental results applying a series of
Schottky formulas to account for the contribution of respective
excited CEF levels:
\begin{equation}
C_{\rm CEF}(T)=\rm{R } \Sigma_i
[(\frac{\Delta_i}{T})/2\cosh(\frac{\Delta_i}{2T})]^2
\end{equation}
being R the gas constant and $\Delta_i$ the energy of respective
levels.

\begin{figure}[tb]
\begin{center}
\includegraphics[width=19pc]{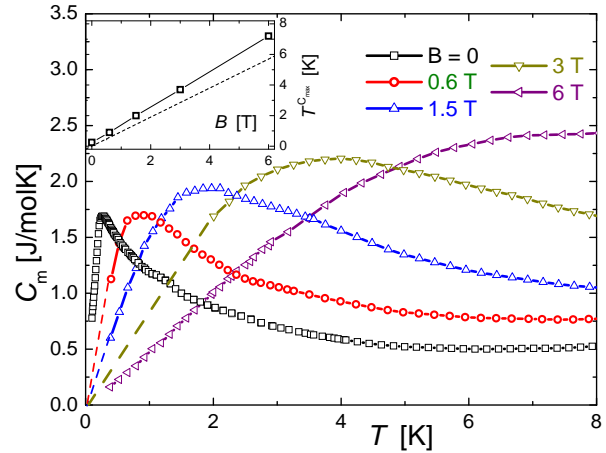}
\end{center}
\caption{Low temperature dependencies of the specific heat of
YbCu$_4$Ni under different magnetic fields 0 - 6\,T. Inset: Field
dependence of the temperature of respective $C_m(T)$ maxima:
$T^{C_{max}}$. Dashed line is the computed dependence for the
$\Gamma_7$ doublet.} \label{FF6}
\end{figure}

According to the result, the first excited doublet is located at
$\Delta_1 = 90$\,K while the second corresponds to the quartet
$\Gamma_8$ at double the energy: $\Delta_2 \approx 210$\,K.
Strictly, the fitting curve included in the inset of
Fig.~\ref{FF5} also contains the actual GS contribution (see
Fig.~\ref{FF6} for more detail) and an improvement obtained by
splitting the $\Gamma_8$ quartet into two doublets at $\Delta'_2 =
190$\,K and $\Delta"_2 = 230$\,K. This further splitting mimics an
eventual level broadening evaluated around 30\,K. This value is in
agreement with the observed $\theta_P = -19$\,K obtained through
an extrapolation from $T\geq 70$\,K where the $\Gamma_8$ quartet
becomes dominant and a Kondo screening effect may be detected. We
remark that specific heat results allow to identify the $\Gamma_8$
quartet as the most excited level because of its larger
degeneracy.

Although the CEF level distribution in YbCu$_4$Ni has some
similarity with that of the isotypic YbCu$_{5-x}$Au$_x$, basically
because $\Gamma_8$ is the upper level \cite{YbCu5-xAux}, the
energy distribution of the levels is quite different. This feature
can be attributed to the different electronic character between
the electron-like Au and the hole-like Ni ligands that presents a
point charge of opposite sign among them that clearly modifies the
Coulomb potential, though without affecting the local structural
symmetry.

\subsection{About the doublet ground state}

The same reason that allows to identify the $\Gamma_8$ quartet as
the most excited level, i.e. its larger degeneracy, impedes to
distinguish between the two doublets: $\Gamma_6$ and $\Gamma_7$ as
ground and first excited level. An analysis based on relative
degeneracies and levels splitting \cite{Erbio} applied to the LLW
tabulations for $J=7/2$ \cite{LLW} gives two equivalent
possibilities for the LLW parameter: $z=0.5$ for a $\Gamma_7$-GS
and $z=-0.3$ for the corresponding $\Gamma_6$-GS. However,
according to previous calculations of the effective moment for
those different pseudospin-1/2 doublets \cite{YbSn3Ru4}
$\mu_{eff}(\Gamma_7)= 2.96\,\mu_B$ nearly coincides with the
experimentally observed value rather than $\mu_{eff}(\Gamma_6)=
2.31\,\mu_B$. This result can be checked by computing the expected
saturation magnetization: $M_{sat}= g_{eff} J_{eff} \mu_B$. For
$g_{eff}(\Gamma_7)=3.43$ and $J_{eff} =1/2$ (for the doublet GS),
one gets $M_{sat}(\Gamma_7)= 1.72\mu_B$ which can be expected from
the $M(B)$ curve at $T=2$\,K included in Fig.~\ref{FF3}a respect
to $M_{sat}(\Gamma_6)= 1.33\mu_B$.

\subsection{Magnetic field effect}

\begin{figure}[tb]
\begin{center}
\includegraphics[width=19pc]{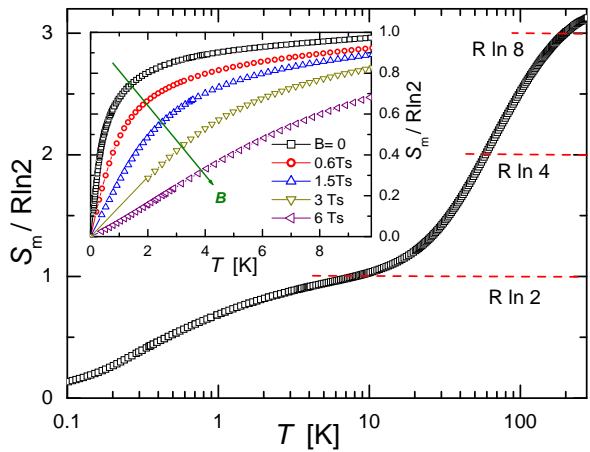}
\end{center}
\caption{The temperature dependence of the magnetic entropy of
YbCu$_4$Ni in a logarithmic temperature scale. The inset shows the
entropy gain with different applied fields, in linear temperature
scale.} \label{FF7}
\end{figure}

The effect of applied magnetic field on the temperature dependence
of specific heat, shed further light on the GS nature through the
Zeeman splitting dependence on $g_{eff}$. This field dependence in
YbCu$_4$Ni is presented in Fig.~\ref{FF6} within the $0.4\leq T
\leq 6$\,K range and fields up to 9\,Tesla. The $C_P(T,B=0)$
maximum ($T_{C_{max}}) \approx 270$\,mK, shifts to higher
temperature proportionally to the field intensity as shown in the
inset of Fig.~\ref{FF6}. Despite the fact that $C_P(T,B=0)$ is
better described as a Kondo-like anomaly \cite{DegrSchot} rather
than a Schottky one (see Fig.~\ref{FF6}), it is expected that the
spin-up and spin-down components split by magnetic field tend to a
Schottky anomaly at high enough field, i.e. $\mu B
>> k_B T$ \cite{Maple}. This criterion, based in the
Zeeman splitting of the GS levels, one can confirm its
$\Gamma_7$-GS nature applying the $T(C_{mx})= 0.42 \times g_{eff}
B (\mu_B/k_B)$ relation \cite{Elsevier} to obtain the dashed line
in the inset of Fig.~\ref{FF6} taking $g_{eff}(\Gamma_7)= 3.43$.
Similar magnetic field effects on $C_m(T,B)$ was reported on cubic
Yb$_{0.24}$Sn$_{0.76}$Ru \cite{YbSn3Ru4}. In this compound,
however, the value of the maximum of $C_m(T)$ is about 35\% larger
than in the stoichiometric compound studied in this work.

\subsection{Entropy}

The magnetic entropy gain ($S_m$) of YbCu$_4$Ni along the full
range of temperature is displayed in Fig.~\ref{FF7} in a
logarithmic temperature scale. It is computed from the magnetic
contribution to the specific heat as $S_m = \int C_m/T dT$.

At room temperature, the entropy slightly exceeds the expected
value for the 8 fold degenerated $J=7/2$ Hund's rule GS of
Yb$^{3+}$ $S_m = R\ln8 = 17.28$\,J/mol\,K. This excess of entropy
is probably due to an under evaluation of the phonon contribution
at high temperature. The flattening of $S_m(T)$ around 10\,K
observed in Fig.~\ref{FF7} corresponds to $S_m$ = R$\ln 2$,
indicating that the doublet GS is well separated from the first
excited CEF level at $\Delta_1 = 85$\,K.

\begin{figure}[tb]
\begin{center}
\includegraphics[width=19pc]{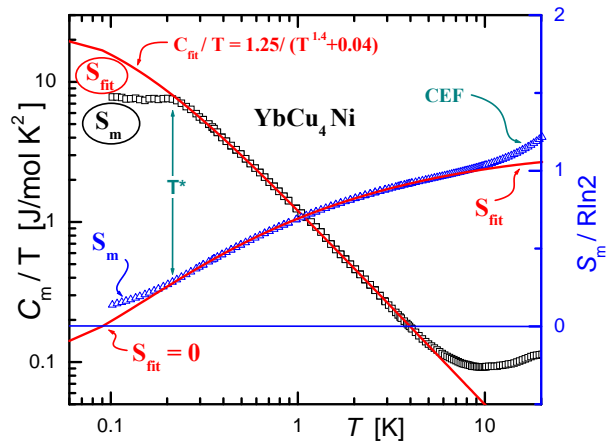}
\end{center}
\caption{Comparison between measured $C_{m}(T)/T$ and fitted
$C_{fit}/T|_{T>T^*}$ (continuous curve) dependencies referred to
the left axis in double logarithmic representation. On the right
axis (linear scale), the temperature dependencies of respective
entropies are referred showing how $S_{fit}\to 0$ at $T>0$. The
value of $S_{fit}(T\to \infty)= $R$\ln2$ is taken as reference for
the doublet GS. CEF indicates the onset of excited crystal field
levels contribution above about 3\,K.} \label{FF8}
\end{figure}

At low temperature, the change of curvature observed below
$T\approx 300$\,mK can be analyzed accounting for the constraints
imposed on the entropy by the Nernst postulate (third law of
thermodynamics) \cite{Pippard}. For such a purpose we compare in
Fig.~\ref{FF8} the temperature dependence of the specific heat
with the entropy. In the double logarithmic representation of the
latter one may appreciate that $C_m(T>T^*)/T$ increases obeying a
power law dependence $\propto 1.25/(T^{1.4}+0.04)$\,J/molK$^2$
which, below $T^*=210$\,mK, transforms into a 'plateau' with
$C_m/T|_{T\to 0} \approx 7.5$\,J/mol\,K$^2$. The power law
dependence, 'plateau' and $T^*$ values are similar to those
reported for YbCu$_{4.3}$Au$_{0.7}$ \cite{YbCu5-xAux}.

The area marked as $S_{m}$ in Fig.~\ref{FF8} represents the
entropy computed from the measured specific heat that corresponds
to the actual entropy released by the system upon cooling. On the
other hand, the area indicated as $S_{fit}$ is extracted as
$S_{fit}=\int C_{fit}/T \times dT$, where $C_{fit}/T=
25/(T^{1.4}+0.04)$\,J/molK$^2$ is the fitted power law dependence
above $T=T^*$. As a mathematical function, $C_{fit}/T$ keeps
growing below $T^*$ as depicted in the figure and therefore the
associated entropy exceeds the physical limit of 'R$\ln2$' for a
doublet GS. Both, $S_{m}$ and $S_{fit}$ entropy trajectories are
compared in the figure (right axis). Since the maximum entropy for
a doublet GS is R$\ln2$, the $S_{fit}$ trajectory (continuous
curve) is drawn in Fig.~\ref{FF8} taking that value as the high
temperature limit. As expected, $S_{fit}$ and $S_{m}$ coincide
within the fitting range (i.e. above $T^*$), with the $S_m(T)$ for
$T>10$\,K being due to the onset of the excited CEF levels
contribution. On the low temperature side, both entropy
trajectories also split but due to different reasons. Since the
physical system obeys the Nernst principle: $S|_{T\to 0} \geq 0$,
$S_{m}(T)$ is constrained to change its trajectory and to point to
$S_{m}=0$ at $T=0$. Otherwise it would become zero as $S_{fit}$
does at finite temperature ($\approx 100$\,mK in this case). Since
that change of trajectory is compelled by thermodynamic
principles, $T^*$ characterizes the temperature at which this sort
of 'entropy bottleneck' occurs \cite{Sereni15}.

\section{Summary}

Based on the parent compound YbCu$_5$, we have replaced one Cu
atom by Ni to obtain the stoichiometric and atomically
well-ordered compound YbCu$_4$Ni with a lattice parameter $a =
6.9429 \AA$ in its cubic structure. At high temperature this
compound exhibits localized magnetic nature and an effective
magnetic moment close to Yb$^{3+}$ according to the Hund's
eight-fold GS. The CEF splits the J=7/2 multiplet into well
defined doublets as GS and first excited state while the expected
quadruplet exhibits a moderate broadening.

At low temperature, this compounds shows clear signs of magnetic
frustration reflected in a power law increase of the specific heat
by decreasing temperature. This diverging tendency is limited by
the Nernst postulate that compels to a change of regime, at a
characteristic temperature $T^* \approx 200$\,mK, which in this
case corresponds to a 'plateau' with very high density of states:
$C_m/T|_{T\to 0}= 7.5$\,J/molK$^2$. Electrical resistivity results
confirm this behavior undergoing a broad maximum around 300\,mK
before to flatten as $T\to 0$. Both features point to the
formation of a Fermi-liquid like GS which does not condensate in a
continuous way as is the case in standard Fermi liquids. This
feature is highlighted by a maximum in the inductive component of
ac-susceptibility at $T\approx T^*$ and a step in the dissipative
component.

Although such a large value of $C_m/T|_{T\to 0}$ was observed in
other very heavy fermion compounds after a change of regime from a
power law $C_m(T)/T$ increase, the discontinuity observed in the
dissipative component of the ac-susceptibility proves that a
change of regime occurs at $T=T^*$ in a continuous drift from a
frustrated GS to a Fermi-liquid one. Applied magnetic field
induces a Zeeman splitting of the doublet ground state, which
progressively tends to a Schottky type anomaly in the specific
heat as the field overcomes the intensity of the ground state
interaction producing the $C_m/T|_{T\to 0}$ 'plateau'.

\vspace{0.5cm}
\section*{Acknowledgements}

MR and IC are partially supported by projects VEGA 1/0956/17 and
VEGA 1/0611/18. AMS thanks the SA-NRF (93549) and the UJ URC/FRC
for financial assistance.


\begin{thebibliography}{00}

\bibitem{YbCu5-xAux} I. Curlik, M. Giovannini, J.G. Sereni, M. Zapotokova, S. Gabani,
M. Reiffers, {\it Extremely high density of magnetic excitations at
$T=0$ in YbCu$_{5-x}$Au$_x$}, Phys. Rev. B {\bf 90} 224409 (2014).

\bibitem{QPT07} H.v. L\"ohneysen, A. Rosch, M. Vojta, P. W\"olfe,
in {\it Fermi-liquid instabilities at magnetic quantum
transitions}, Rew. Mod. Phys. {\bf 79} (2007) 1015.

\bibitem{Stewart01} G.R. Stewart, in {\it Non-Fermi-liquid behavior
in d- and f-electron metals}, Rew. Mod. Phys. {\bf 73} (2001) 797.

\bibitem{JLTP18} J.G. Sereni, in {\it Entropy constraints in the
ground state formation of magnetically frustrated systems}, J. Low
Temp. Phys. {\bf 190} (2018) 1-19, and references therein.

\bibitem{YbPt2Sn} T. Gruner, D. Jang, A. Steppke, M. Brando, F.
Ritter, C. Krellner, C. Geibel, in {\it Unusual weak magnetic
exchange in YbPt$_2$Sn and YbPt$_2$In}, J. Phys.: Condens. Matter
{\bf 26} 485002 (2014).

\bibitem{Ramirez06} R. Moessner and A.P. Ramirez, in {\it
Geometrical Frustration}; Physics Today, February 2006, p.24.

\bibitem{Sereni15} J.G. Sereni, in {\it Entropy Bottlenecks at $T\to 0$ in Ce-Lattice and Related
Compounds}, J Low Temp Phys {\bf 179} (2015) 126–137.

\bibitem{PrInAg2} A. Yatskar, W.P. Beyermann, R. Movshovich, P.C. Canfield ; Phys. Rev. Lett. {\bf 77} (1996) 3637.

\bibitem{Caretta} P. Carretta, R. Pasero, M. Giovannini, C. Baines, Phys.Rev.B,
{\bf 79}, R020401 (2009).


\bibitem{GrunerNat} D. Jang, T. Gruner, A. Steppke, K. Mistsumoto, C. Geibel, M.
Brando, in {\it Large magnetocaloric effect and adiabatic
demagnetization refrigeration with YbPt$_2$Sn}, Nature
Communications, ncomms9680 (2015).

\bibitem{Gegenw} Y. Tokiwa, B. Piening, H.S. Jeevan, S.L. Bud'ko, P.C. Canfield, P.
Gegenwart,in {\it Super-heavy electron material as metallic
refrigerant for adiabatic demagnetization cooling}, Sci. Adv. 2
(2016) e1600835.

\bibitem{YbCo2Zn20} M.S. Torikachvili, S. Jia, E.D. Mun, S.T. Hannahs, R.C. Black,
W.K. Neils, D. Martien, S.L. Bud'ko, P.C. Canfield, in {\it Six
closely related related YbT$_2$Zn$_{20}$ heavy fermion compounds
with large local moment degeneracy}, PNAS {\bf 104} (2007) 9960.

\bibitem{YbBiPt} Z. Fisk, P.C. Canfield, W.P. Beyermann, J.D. Thompson,
M.F. Hundley, H.R. Ott, E. Felder, M. B. Maple, M.A. Lopez de la
Torre, P. Visani, C. L. Seamanet; . Phys. Rev. Lett. {\bf 67}
(1991) 3310.

\bibitem{YbCu4Au} M. Galli, E. Bauer, St. Berger, Ch. Dusek, M. Della Mea, H. Michor, D.
Kaczorowski, E.W. Scheidt, F. Marabelli, in {\it Evolution of
ground state properties of YbCu$_{5-x}$Au$_x$}, Physica B {\bf 312
\& 313} (2002) 489–491.

\bibitem{Giova15} M. Giovannini, I. Cûrlík, F. Gastaldo, M. Reiffers,
J.G. Sereni, {\it The role of crystal chemistry in
YbCu$_{5-x}$Au$_x$}, Journal of Alloys and Compounds {\bf 627}
(2015) 20–24.

\bibitem{Fullprof} J. Rodriguez-Carvajal, Physica B {\bf 192} (1993) 55.

\bibitem{12Curlik} I. Curlik, M. Reiffers, M.
Giovannini, in {\it Study of Magnetic contribution to the Heat
Capacity of YbCu4Ni}, Acta Phys. Pol. A {\bf 122} (2012) 3-5.

\bibitem{rhoActa} I. Curlik, S. Matosova, S. Ilkovic, M. Reiffers, M.
Giovannini, in {\it Transport and Magnetic Properties of
YbCu$_4$Ni}, Acta Phys. Pol. A {\bf 122} (2012) 6-8.

\bibitem{Erbio} J.G. Sereni, in {\it Crystal Field Effects in
Metals and Alloys}, Ed. A. Furrer, Plenum Press, N.Y., 1977, 309.

\bibitem{DegrSchot} H.U. Desgrages, K.D. Schotte, in {\it Specific heat of the Kondo
model}, Phys. Lett. A {\bf 91} (1982) 240.

\bibitem{Maple} M.B. Maple, in {\it Superconductivity: a probe for
the magnetic state of local moments in metals}, Appl. Phys. {\bf
9} (1976) 179-204.


\bibitem{LLW} W.E. Lea, M.J.M. Leask, W.P. Wolf, {\it The raising of angular momentum
degeneracy of f-electron terms by cubic crystal fields}, J. Phys.
Chem. Solids {\bf 23} (1962) 1381-1405.


\bibitem{YbSn3Ru4} T. Klimczuk, C.H. Wang, J.M. Lawrence, Q. Xu, T.
Durakiewicz, F. Ronning, A. Llobet, F. Trouw, N. Kurita, Y.
Tokiwa, Han-oh Lee, C.H. Booth, J.S. Gardner, E. D. Bauer, J.J.
Joyce, H.W. Zandbergen, R. Movshovich, R.J. Cava, J.D. Thompson,
in {\it Crystal fields, disorder, and antiferromagnetic
short-range order in Yb$_{0.24}$Sn$_{0.76}$Ru}, Phys. Rev. B {\bf
84} (2011) 075152.


\bibitem{Elsevier} J.G. Sereni, in: {\it Magnetic Systems: Specific Heat}, Saleem Hashmi (editor-in-chief),
Materials Science and Materials Engineering. Oxford: Elsevier;
2016. pp. 1-13; ISBN: 978-0-12-803581-8

\bibitem{Pippard} A.B. Pippard, in {\it Elements of classical
Thermodynamics}, University Press, Cambridge, 1964.

\end{thebibliography}
\end{document}